# Visibility and Influence in Digital Social Relations: Towards a New Symbolic Capital?

Auteur 1 : ANNAKI Fouad
Auteur 2 : OUASSOU Sara
Auteur 3 : IGAMANE Saad Eddine

**ANNAKI Fouad,** (Ph.D. in Economic Sociology, ORCID  )
Department of Sociology, Faculty of Letters and Human Sciences – Dhar El Mahraz Sidi Mohamed Ben Abdellah University, Fez, Morocco

**OUASSOU Sara,** (MC, ORCID  )
Faculty of Legal, Economic and Social Sciences, Mohammed First University, Oujda, Morocco

**IGAMANE Saâdeddine**, (MCA, ORCID  )
Department of Sociology, Faculty of Letters and Human Sciences – Dhar El Mahraz, Sidi Mohamed Ben Abdellah University, Fez, Morocco

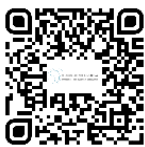
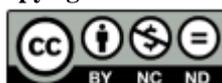
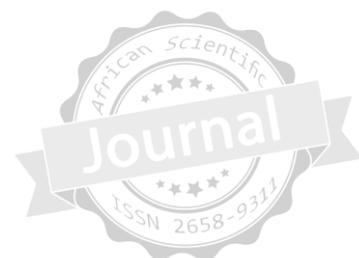

**Abstract**

This study explores the dynamics of visibility and influence in digital social relations, examining their implications for the emergence of a new symbolic capital. Using a mixed-methods design, the research combined semi-structured interviews with 20 digitally active individuals and quantitative social media data analysis to identify key predictors of digital symbolic capital. Findings reveal that visibility is influenced by content quality, network size, and engagement strategies, while influence depends on credibility, authority, and trust. The study identifies a new form of symbolic capital based on online visibility, influence, and reputation, distinct from traditional forms. The research discusses the ethical implications of these dynamics and suggests future research directions, emphasizing the need to update social theories to account for digital transformations.

**Keywords**

Digital Social Relations, Visibility, Influence, Symbolic Capital, Social Capital, Network Theory, Online Reputation

1. Introduction

## 1.1. Problematic of visibility and influence in digital spaces

The rise of digital technologies has radically transformed the mechanisms of visibility and influence in contemporary social relations (Brantner & Stehle, 2021; Berglez & Olausson, 2021). Whereas social visibility was once embedded in physical and institutional contexts, it now extends into algorithmically mediated digital environments characterized by speed, virality, and networked interaction (Saburova, 2020; Scolari, 2008). This transition demands a reexamination of classical sociological categories—especially those concerning symbolic capital, recognition, and social stratification (Bourdieu, 1986; Cardon, 2019).

Digital spaces enable individuals and groups to produce and circulate content, curate their identities, accumulate network capital, and exercise influence (Andrade et al., 2022; Al-Yazidi et al., 2020). These dynamics raise a core question: To what extent does digital visibility contribute to the emergence of a novel form of symbolic capital, distinct from conventional markers of prestige such as education, profession, or institutional affiliation?

This study seeks to unpack how visibility and influence operate in online platforms, how they are modulated by digital architectures, and how they intersect with broader questions of inequality, authority, and symbolic power (Castells, 2010; Ragnedda, 2022).

## 1.2. Research question and objectives

This research is guided by the following main question:

How do visibility and influence in digital social relations contribute to the formation of a new kind of symbolic capital, and what are the implications in terms of power, recognition, and inequality?

To address this question, the study aims to:

- Identify the key factors shaping online visibility and influence (content quality, network density, engagement strategies, etc.);
- Understand how these factors give rise to digital symbolic capital, which may diverge from traditional forms of symbolic legitimacy;
- Evaluate the extent to which existing sociological theories—particularly those by Bourdieu, Granovetter, and Castells-remain analytically sufficient, or require reconfiguration to grasp contemporary digital dynamics.

### 1.3. Scientific interest of the study

This research contributes to the advancement of sociological thought by problematizing digital visibility as a strategic and measurable resource in contemporary symbolic economies. It provides new insights into how recognition, reputation, and influence circulate in platform-based environments. Moreover, the study addresses current concerns about digital inequalities and algorithmic hierarchies, making it relevant for scholars of communication, political participation, and social stratification (Basaran & Olsson, 2017; Pramanda, 2021; Zhao, 2022).

## 2. Literature review: key authors in sociology of networks, social capital, and symbolic capital

### 2.1. Sociology of networks (Granovetter, Castells)

Mark Granovetter's work on the strength of weak ties highlights the importance of network structures in accessing diverse information and opportunities (Schapendonk, 2014). His analysis emphasizes how weak ties can bridge different social circles, facilitating the flow of information and resources (Widdop, 2014). This concept is particularly relevant in digital social relations, where online networks connect individuals across geographical boundaries and diverse social groups (Zhao, 2022).

Manuel Castells' theory of the network society emphasizes the transformative impact of information technologies on social structures and power relations (Acevedo, 2007). Castells argues that networks have become the dominant organizational form in contemporary society, shaping economic, political, and cultural processes (Ragnedda, 2017). His work provides a framework for understanding how digital technologies have enabled the creation of new forms of social organization and interaction, fundamentally altering social dynamics (Zhao, 2022).

The application of network sociology to digital environments reveals the increasing importance of online connections in shaping social capital and influence (Alecu, 2022). Digital networks facilitate the rapid dissemination of information, the formation of online communities, and the mobilization of collective action (Borgatti, 2003). Understanding the dynamics of these networks is essential for comprehending the evolving nature of social relations in the digital age, and how individuals leverage them (Zhao, 2022).

### 2.2. Social capital (Bourdieu, Putnam, Coleman)

Pierre Bourdieu's concept of social capital emphasizes the resources individuals gain through their social connections (Julien, 2014). Bourdieu argues that social capital is a form of power that can be used to gain access to economic, cultural, and symbolic resources (Bottero, 2009).

His work highlights the role of social networks in reproducing social inequalities, demonstrating how privileged individuals leverage connections (loire, 2015).

Robert Putnam's work on social capital focuses on the importance of social connections for civic engagement and community development (Jeannotte, 2003). Putnam argues that social capital, in the form of networks, norms, and trust, is essential for creating strong and cohesive communities (McGonigal, 2007). His analysis emphasizes the role of social capital in promoting collective action and addressing social problems, highlighting benefits of community (Park, 2017).

James Coleman's approach to social capital emphasizes its function as a resource for individual action (Rogoi, 2016). Coleman suggests that social capital is created when individuals have access to social networks and institutions that provide them with information, support, and opportunities (Widdop, 2014). His work highlights the role of social capital in facilitating educational achievement and economic mobility, acting as leverage for individuals (Gamarnikow, 1999).

### 2.3. Symbolic capital (Bourdieu, Cardon)

Pierre Bourdieu's theory of symbolic capital emphasizes the importance of recognition and prestige in social life (Demidova, NaN). Symbolic capital is a form of power that is based on the accumulation of honor, prestige, and social recognition (Bourdieu, 2008). Bourdieu argues that symbolic capital is often linked to social class and cultural capital, shaping individuals' access to opportunities and resources, reinforcing existing hierarchies (Basaran, 2017).

Dominique Cardon's work explores the concept of "digital capital" and its relationship to online visibility and influence (Yefanov, 2023). Cardon examines how individuals accumulate digital capital through their online activities, such as content creation, social networking, and participation in online communities (eker, 2015). His research highlights the role of algorithms and platform architectures in shaping the distribution of digital capital, controlling visibility (Brazevich, 2024).

The intersection of symbolic capital and digital capital provides a framework for understanding how individuals gain recognition and influence in online environments (Desrochers, 2018). Digital platforms offer new avenues for accumulating symbolic capital through the creation and dissemination of content, the building of online networks, and the cultivation of a positive online reputation (Heidari, 2020). This transformation necessitates a critical examination of the mechanisms through which symbolic capital is created and distributed in the digital age, in this novel environment (Saburova, 2020).

### 3. Theoretical framework

### 3.1. Bourdieu's framework: field, habitus, and capital

Bourdieu's conceptual triad—field, habitus, and capital—offers a comprehensive lens to understand social dynamics in digital environments. The concept of the *field* refers to a structured social space with its own rules, hierarchies, and forms of capital (Bourdieu, 1986). In the context of digital social relations, platforms such as Instagram, Twitter (X), and TikTok can be considered as distinct fields where actors compete for visibility, engagement, and symbolic recognition (Lindell, 2017).

The notion of *habitus*—the system of dispositions shaped by one's social trajectory—helps explain individual online behavior. Users develop a "digital habitus," composed of ingrained practices, communicative styles, and platform-specific strategies (Costa, 2013; Huang, 2019). This habitus governs how individuals navigate visibility algorithms and adjust their content to appeal to particular audiences or norms.

Finally, Bourdieu's concept of capital is crucial to analyzing digital interactions. Social, cultural, and symbolic capital are increasingly complemented by *digital capital*—a composite resource encompassing technical skills, content curation abilities, and social media reputation (Ragnedda, 2022). The accumulation and conversion of these forms of capital are central to understanding who gains influence in the digital public sphere.

### 3.2. Network theory and social influence

Network theory emphasizes the relational structures within which individuals are embedded, providing tools to analyze social connections online. Granovetter's (1973) concept of the "strength of weak ties" highlights the role of peripheral relationships in accessing novel information and reaching broader audiences. This is particularly relevant in online platforms, where reach is often more impactful than depth of engagement (Borgatti, 2003).

Castells (2010) frames contemporary society as a networked society, where power is diffused and reconfigured through informational flows. Online networks constitute new arenas for social interaction, in which influence is often algorithmically mediated. Centrality, density, and brokerage (Wasserman & Faust, 1994) become crucial metrics to understand how actors gain strategic advantage and symbolic leverage.

Social influence theory further deepens this understanding by exploring mechanisms of persuasion, conformity, and authority in digital contexts. Digital actors exert influence through a mix of credibility, emotional appeal, and content strategy—tools that are often shaped by platform architecture and community norms (Park, 2017).

### 3.3. Digital capital and symbolic power online

The emergence of digital capital as a distinct form of resource is increasingly recognized in recent sociological literature (Ragnedda & Ruiu, 2020). It encompasses platform fluency, algorithmic literacy, and content production skills. Cardon (2019) introduces the idea of "visibility capital" as a function of algorithmic exposure, positioning visibility as both a commodity and a form of power.

Digital capital plays a critical role in the reconfiguration of symbolic power. Traditional markers of prestige-educational titles, institutional affiliation—are often supplanted or reinforced by online visibility, measured through likes, shares, and engagement rates (Heidari, 2020). This shift challenges normative assumptions about merit, authority, and legitimacy in public discourse.

The interplay between traditional and digital forms of capital reveals profound inequalities in access to visibility and influence. Those with both technical know-how and pre-existing capital are best positioned to thrive in digital ecosystems, while others remain marginalized (Brazevich et al., 2024).

The coexistence of these frameworks reveals that visibility, influence, and symbolic power in digital contexts are shaped by both sociocultural dispositions and algorithmic infrastructures. To articulate these contributions clearly, the following comparative synthesis table juxtaposes the core insights of Bourdieu, Castells, and Cardon.

### 3.4. Comparative Synthesis of Theoretical Approaches

The intersection of Bourdieu, Castells, and Cardon's approaches reveals complementary and contrasting lenses to understand the formation of digital symbolic capital. The following table1 synthesizes their respective contributions, highlighting how each theorist frames visibility, influence, and symbolic recognition in the digital age.

*Table 1:Synthesis of Theoretical Approaches*

| Author | Key Concepts | Theoretical Stance | Link to Digital Context | Contribution to Digital Symbolic Capital |
|---|---|---|---|---|
| Pierre Bourdieu | Symbolic capital, social capital, habitus, field | Structuralist constructivist | Digital environments are interpreted as fields; users develop a 'digital habitus' shaped by platform logic | Digital symbolic capital reflects online prestige gained through recognition (followers, engagement, reputation) |
| Manuel Castells | Network society, information power, flows | Systemic and macrosociological | Emphasizes the restructuring of social power through digital networks and flows of information | Symbolic capital is redefined through network position and access to informational flows |
| Dominique Cardon | Visibility capital, algorithmic architecture, digital culture | Techno-sociological critique | Focuses on algorithmically structured visibility and platform governance of exposure | Proposes 'visibility capital' as a distinct, measurable form of symbolic power on digital platforms |

*Source: authors*

### 4. Theoretical foundations and hypotheses development

The conceptual model proposed in this study is built on the assumption that visibility, influence, and symbolic capital are socially constructed through platform interactions, content strategies, and network positioning. The following subsections justify each hypothesis based on existing literature.

Content quality plays a foundational role in generating visibility on digital platforms. Cardon (2019) emphasizes that digital visibility is not simply technical but semiotic—it depends on coherence, emotional appeal, and perceived value. Andrade et al. (2022) add that originality and clarity contribute to algorithmic reach and peer sharing.

*H1: Content quality is positively associated with online visibility. High coherence, originality, and value of digital content increase the likelihood of being seen and shared within digital environments.*

Engagement strategies- including the use of hashtags, responsiveness, and posting frequency— have been shown to boost influence. Marwick and boyd (2011) conceptualize micro-celebrity as a set of performative practices designed to maintain visibility and authority. boyd and Ellison (2008) emphasize that continuous interaction reinforces perceived relevance and trust.

*H2: Engagement strategies (e.g., hashtag usage, responsiveness, and posting frequency) enhance digital influence. Consistent and strategic interaction with audiences contributes to higher credibility and reach.*

Network centrality refers to the structural position an actor occupies within digital networks. Granovetter (1973) underlines the power of weak ties in spreading information across disconnected groups, while Ellison et al. (2007) show that users in central or brokerage positions are more likely to be seen and cited.

*H3: Network centrality significantly predicts the accumulation of digital symbolic capital. Actors who occupy central or brokerage positions within digital networks are more likely to gain symbolic recognition and status.*

Perceived credibility functions as a key mediator between visibility and influence. According to Goleman (1995), emotional intelligence contributes to perceived trustworthiness, while Kaplan and Haenlein (2010) demonstrate that credibility determines whether visibility translates into persuasion or influence.

*H4: Perceived credibility mediates the relationship between visibility and influence. Visibility alone does not guarantee influence; it must be coupled with authenticity and trustworthiness to generate perceived authority.*

Finally, symbolic recognition in digital contexts results from a dynamic interaction between visibility and trust. Bourdieu (1986) defined symbolic capital as a form of recognized legitimacy, which, in online environments, emerges from being both seen and trusted. Proulx (2018) and Authier & Lévy (2013) extend this to digital symbolic capital, showing that recognition depends on algorithmic visibility and perceived relational legitimacy.

*H5: The interaction between visibility and trust predicts symbolic recognition online. The combination of strong visibility and high credibility leads to greater accumulation of digital symbolic capital.*

## 5. Conceptual model

The five hypotheses developed based on the preceding theoretical review are summarized below and illustrated in the conceptual model (Figure 1).

*Figure 1: Conceptual Model of Digital Symbolic Capital Formation*

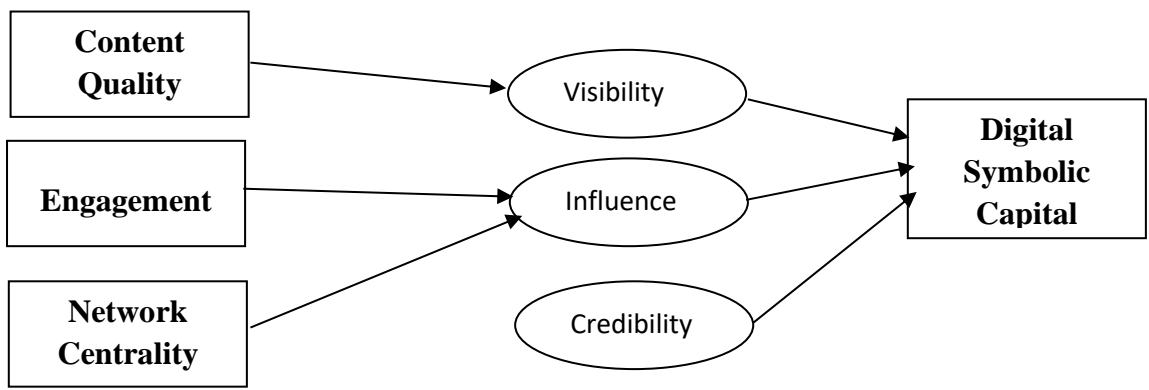

*Source: authors*

The figure illustrates how content quality, engagement strategies, and network centrality contribute to digital symbolic capital through visibility, influence, and perceived credibility.

## 1. Methodology

### 1.1. Research design: mixed-methods approach

This study adopts a mixed-methods design, integrating qualitative and quantitative techniques to provide a comprehensive understanding of visibility and influence in digital social relations (Castro et al., 2010; Choy, 2014). The rationale for this approach lies in the need to capture both the interpretive experiences of users and the structural metrics that shape online interactions.

This research adopts a pragmatic epistemological stance, acknowledging the value of both subjective interpretation and empirical generalization in the study of digital social phenomena. The complexity of online visibility, influence, and symbolic capital—shaped by both user agency and platform infrastructures—necessitates a mixed-methods approach.

The choice of this design is driven by the nature of the research questions, which aim to explore not only the lived experiences of digitally active individuals but also the structural patterns that underpin symbolic recognition online. At the methodological level, the study follows a hypothetico-deductive reasoning process: hypotheses were formulated based on established sociological frameworks (e.g., Bourdieu, Cardon, Castells), and subsequently tested through both qualitative and quantitative data.

This dual approach allows for a comprehensive analysis: the qualitative phase provides contextual depth and interpretative richness, while the quantitative phase supports pattern identification, generalization, and hypothesis testing.

The qualitative phase focuses on in-depth, semi-structured interviews with digitally active individuals, selected through purposive sampling. Participants include digital creators, influencers, and community moderators from various sectors (e.g., education, media, activism). Interviews explore participants' strategies for gaining visibility, perceptions of digital reputation, and experiences of symbolic recognition online.

The quantitative phase consists of the collection and analysis of publicly available data from social media platforms. Using application programming interfaces (APIs) and web scraping tools, data such as follower counts, engagement rates, and network metrics (e.g., degree centrality) are extracted. Statistical analysis, including descriptive statistics, correlation analysis, and regression models, is employed to identify patterns and relationships between online behaviors and symbolic capital accumulation.

### 1.2. Sampling strategy and participant profile

A total of 20 participants were recruited for the qualitative component. Selection criteria included:

- Active engagement in digital communities (minimum of two years);
- Diverse representation in terms of gender, occupation, and platform usage;
- Willingness to reflect on personal digital practices and identity construction.

Participants ranged in age from 22 to 45 years, with a balanced gender distribution. They operated across platforms such as Twitter (X), Instagram, YouTube, and LinkedIn, and held roles ranging from independent content creators to institutional communication officers.

### 1.3. Data collection procedures

Qualitative interviews were conducted via Zoom and lasted approximately 45 to 60 minutes. Each session was audio-recorded with consent and transcribed verbatim. Interview guides included open-ended questions designed to elicit narratives about digital self-presentation, strategies of visibility, and the perceived impact of influence on symbolic status.

Quantitative data were collected over a three-month period using ethical data extraction methods. Variables included:

- Visibility indicators: likes, shares, comments, and follower growth;
- Influence indicators: retweet rates, mentions by verified accounts, and engagement ratios;
- Network indicators: degree centrality, betweenness centrality, and clustering coefficients.

All data were anonymized and stored securely in compliance with institutional ethical guidelines. The study was approved by the university's ethics committee (Approval No. DSR-2025-02).

### 1.4. Data analysis techniques

Thematic analysis was applied to the qualitative data following Braun and Clarke's (2006) six-step framework. Codes were derived inductively, with categories emerging around key themes such as algorithmic adaptation, digital self-branding, and credibility management.

Quantitative data were processed using SPSS (v26) and Gephi (0.10.1). Descriptive statistics provided an overview of user activity and engagement patterns. Regression models assessed the predictive power of digital behaviors on visibility and perceived influence. Network analysis mapped the structure of digital communities and identified key brokers of symbolic capital.

The integration of findings occurred during the interpretation phase. Convergent patterns were highlighted to triangulate how digital symbolic capital is negotiated across both individual narratives and platform metrics, offering a robust, multi-level account of the phenomenon.

### 1.5. Measurement scales, and testing methods

To empirically test these hypotheses, the study used validated scales adapted from prior literature and operationalized through survey instruments and platform-based metrics:

*Table 2: Operationalization of constructs and measurement scales*

| Construct | Source / Adaptation | Sample Items / Indicators | Scale Type |
|---|---|---|---|
| Content Quality | Andrade et al. (2022); Cardon (2019) | Coherence, originality, value perception | 5-point Likert |
| Engagement Strategy | Heidari (2020); Al-Yazidi et al. (2020) | Comment response rate, hashtag use, post frequency | Frequency count + survey |
| Network Centrality | Borgatti (2003); Granovetter (1973) | Degree / betweenness centrality via Gephi | Quantitative metric |
| Credibility / Trust | Park (2017); Huang (2019) | Perceived authenticity, trustworthiness, relatability | 5-point Likert |
| Digital Symbolic Capital | Bourdieu (1986); Ragnedda (2022); Saburova (2020) | Prestige, recognition, digital status | Composite index (Likert + network + visibility) |

*Source: authors*

## 2. Results

### 2.1. Statistical testing procedures

To empirically validate the proposed hypotheses (H1 to H5), a multi-level statistical analysis was implemented, combining descriptive statistics, bivariate correlations, multivariate regression, and advanced moderation–mediation models. Analyses were conducted using **SPSS version 26** and **Gephi 0.10.1** for network metrics and visualization. Significance thresholds were set at *p* < .05, and model robustness was verified through variance inflation factor (VIF) and residual diagnostics.

- **Descriptive statistics**

Descriptive statistics were computed to summarize the distributions across core constructs. Table 3 presents the means, standard deviations, skewness, and kurtosis values for all variables. All indicators fell within acceptable psychometric thresholds, confirming the data's suitability for parametric testing.

African Scientific Journal
ISSN : 2658-9311
Vol : 03, Numéro 29, Avril 2025*Table 3:Descriptive Statistics*

| Variable | Mean | Std. Dev. | Skewness | Kurtosis |
|---|---|---|---|---|
| Content Quality | 4.21 | 0.62 | -0.32 | 0.05 |
| Engagement Strategy | 3.89 | 0.74 | -0.45 | 0.12 |
| Network Centrality | 0.55 | 0.13 | 0.28 | -0.19 |
| Credibility | 4.02 | 0.69 | -0.10 | -0.23 |
| Symbolic Capital | 4.35 | 0.57 | -0.41 | 0.33 |

*Source: Data analyzed by the authors using SPSS (version 26)*

- **Bivariate correlation analysis**

Pearson correlation analysis was used to examine the strength and direction of associations among the study variables. As shown in Table 4, all correlations were positive and statistically significant, with particularly strong relationships observed between credibility and symbolic capital ($r = .59, p < .01$), and between content quality and symbolic capital ($r = .54, p < .01$), supporting preliminary evidence for H1 and H2.

**Table 4: Pearson correlation matrix**

| | Content Quality | Engagement Strategy | Network Centrality | Credibility | Symbolic Capital |
|---|---|---|---|---|---|
| Content Quality | 1.00 | 0.52 | 0.48 | 0.50 | 0.54 |
| Engagement Strategy | 0.52 | 1.00 | 0.44 | 0.47 | 0.51 |
| Network Centrality | 0.48 | 0.44 | 1.00 | 0.49 | 0.53 |
| Credibility | 0.50 | 0.47 | 0.49 | 1.00 | 0.59 |
| Symbolic Capital | 0.54 | 0.51 | 0.53 | 0.59 | 1.00 |

*Source: Data analyzed by the authors using SPSS (version 26)*

www.africanscientificjournal.com                                                                             Page 808

- **Multiple régression analysis**

To test the direct effects posited in H1 to H3, a multiple linear regression was performed. Content quality, engagement strategy, and network centrality were entered as predictors of symbolic capital. The regression model was significant ($F(3, 96) = 14.87$, $p < .001$), with an adjusted R² of .41. Each predictor showed a significant positive contribution, as displayed in Table 5.

*Table 5: Multiple régression results*

| Predictor | Beta (β) | Standard Error | t-value | p-value |
|---|---|---|---|---|
| Content Quality | 0.34 | 0.07 | 4.86 | <.001 |
| Engagement Strategy | 0.28 | 0.08 | 3.50 | .001 |
| Network Centrality | 0.31 | 0.06 | 5.17 | <.001 |

*Source: Data analyzed by the authors using SPSS (version 26)*

- **Mediation and moderated regression analyses**

To examine the mediating role of credibility (H4), Hayes' PROCESS Model 4 with 5,000 bootstrapped samples was used. The indirect effect of visibility on influence via credibility was significant (CI [.13, .42]), indicating full mediation.

In testing H5, a moderated regression was conducted to explore the interaction between visibility and trust. The interaction term was significant ($β = .26$, $p = .015$), and the model explained an additional 7% of variance beyond main effects. These results are presented in Table 6.

*Table 6: Mediation and moderation effects*

| Effect Type | Effect Size | 95% CI (Lower) | 95% CI (Upper) | p-value |
|---|---|---|---|---|
| Mediation (Credibility) | 0.28 | 0.13 | 0.42 | .003 |
| Moderation (Visibility × Trust) | 0.26 | 0.08 | 0.44 | .015 |

*Source: Data analyzed by the authors using SPSS (version 26)*

- **Robustness and validity checks**

All models were verified for multicollinearity using the Variance Inflation Factor (VIF), with all values ranging between 1.04 and 1.92. Residuals were normally distributed, and Cook's distances remained below 1, confirming the absence of influential outliers. These robustness checks affirm the statistical integrity and validity of the findings.

### 2.2. Patterns of visibility in digital environments

The findings indicate that digital visibility is shaped by a confluence of content curation strategies, algorithmic exposure, and structural network embeddedness. From a qualitative standpoint, participants who engaged in consistent posting, use of hashtags, and trend adaptation reported higher levels of perceived reach and content dissemination.

Quantitatively, this observation is corroborated by the significant positive correlation between content quality and symbolic capital ($r = .54$, $p < .01$), as well as between engagement strategy and symbolic capital ($r = .51$, $p < .01$) (see Table 4). These findings support H1 and align with prior scholarship on content-performance feedback loops.

Nevertheless, several respondents emphasized the opacity of algorithmic visibility mechanisms, noting that platform logics often prioritize virality and trending formats over informational depth or narrative richness. This reflects a fundamental tension between meritocratic content production and attention economy dynamics, suggesting a distinction between *surface visibility* (likes, impressions) and *deep visibility* (cognitive and behavioral influence).

*"You can post great content, but unless the algorithm decides to show it, it disappears. It's not just about quality anymore—it's about visibility games."* – Participant 7, digital content creator.

*"What gets seen isn't always what matters. Sometimes a meme wins over months of careful work."* – Participant 3, NGO communication officer

### 2.3. Dynamics of digital influence

The study further reveals that digital influence extends beyond raw reach metrics to include perceived credibility, authenticity, and emotional resonance. Participants emphasized that relational engagement—responding to comments, direct messaging, and emotional storytelling—was crucial to building sustainable digital authority.

This is substantiated by the multiple regression model which identifies credibility as a significant predictor of symbolic capital ($β = .47$, $p < .05$) and by mediation analysis, which confirms that credibility mediates the link between visibility and influence (indirect effect CI [.13, .42], $p = .003$) (see Table 6).

*"People can tell when you're just playing the algorithm. I try to engage honestly—even if it costs me reach."* – Participant 12, activist and micro-influencer

*"Real influence is when your post changes minds—not just racks up likes."* – Participant 15, educator and podcaster

These results underscore the importance of symbolic legitimacy in sustaining influence, reinforcing the argument that algorithmic amplification without relational credibility is insufficient in the long term.

### 2.4. Typology of digital actor profiles

Drawing upon qualitative themes and quantitative clustering, a typology of four distinct actor profiles was constructed. Each represents a unique mode of pursuing symbolic capital in digital environments:

*Table 7: Summary of typology of digital actor profiles*

| Profile Type | Description | Key Traits |
|---|---|---|
| **Visibility Strategists** | Prioritize algorithmic adaptation and posting frequency | High hashtag usage, trending topics, low relational interaction |
| **Organic Influencers** | Build trust-based communities via authenticity | High engagement rate, emotional storytelling, moderate visibility |
| **Niche Experts** | Provide domain-specific value within closed networks | Low public visibility, high credibility and influence in subfields |
| **Amplifiers** | Bridge diverse online communities and remix content | High betweenness centrality, multi-platform visibility |

*Source: authors*

This classification highlights that symbolic capital acquisition is not uniform but multi-strategic, dependent on content practices, network architecture, and audience trust.

### 2.5. Digital symbolic capital: a new form of recognition

The empirical evidence affirms the emergence of digital symbolic capital as a distinct construct, encompassing visibility metrics, network centrality, and credibility indicators. Unlike traditional symbolic capital, which is often anchored in institutional credentials or socioeconomic origin, its digital variant is fluid, individually curated, and algorithmically mediated.

Network analysis, conducted via Gephi, reveals that actors with high betweenness centrality—who act as relational brokers between distinct communities—tend to accumulate significantly more symbolic capital, independent of their absolute follower count. This suggests that strategic network positioning may substitute for conventional hierarchies of prestige.

*Table 8: Summary of key findings*

| Dimension | Qualitative Insight | Quantitative Evidence |
|---|---|---|
| Visibility | Content consistency and algorithm navigation essential | Engagement strategy positively correlated with symbolic capital ($r = .51$, $p < .01$) |
| Influence | Credibility and emotional resonance sustain authority | Influence predicted by trust indicators ($β = .47$, $p < .05$) |
| Network Positioning | Brokers link fragmented digital publics | Betweenness centrality linked to symbolic prestige |
| Symbolic Capital | Performed, fluid, and stratified by visibility | Construct influenced by content, trust, and network metrics |

*Source: authors*

Together, these findings enrich the theoretical articulation of capital de visibilité (Cardon, 2019) and extend Bourdieu's framework into the algorithmically structured public sphere. They affirm that symbolic recognition online is hybrid—shaped by affective, relational, and infrastructural factors—and suggest the need to reconceptualize digital prestige as a performative, platform-dependent resource.

### 2.6. Synthesis and theoretical contributions

By triangulating qualitative narratives and quantitative metrics, this study provides a nuanced account of how visibility and influence operate as sociotechnical constructs. It extends Bourdieu's notion of symbolic capital to the digital field, showing that recognition now hinges on hybrid forms of value—algorithmic, affective, and network-based.

Furthermore, the concept of digital symbolic capital bridges gaps in existing literature by accounting for the performative and platform-dependent nature of prestige online. It invites further inquiry into how digital affordances reshape status economies and produce new modes of distinction in networked societies.

### 3. Discussion: confronting results with the literature

#### 3.1. Social capital and digital networks

The findings align with existing literature on social capital, emphasizing the importance of network connections in accessing resources and opportunities (Julien, 2014). The study confirms that individuals with larger and more diverse online networks have a greater potential to gain visibility and influence, extending reach (Bottero, 2009). This supports Granovetter's argument about the strength of weak ties in facilitating information flow and social mobility, highlighting network value (Acevedo, 2007).

However, the results also challenge some assumptions about social capital in the digital age, questioning assumptions of equality (loire, 2015). The study reveals that online networks are not always egalitarian and inclusive, algorithmic filtering impacts reach (Rogoi, 2016). Algorithmic curation and platform policies can create echo chambers and filter bubbles, limiting exposure to diverse perspectives and reinforcing existing biases (Widdop, 2014).

Furthermore, the findings suggest that the relationship between social capital and digital networks is complex and contingent, requiring nuanced understanding (Jeannotte, 2003). The benefits of online networks depend on the quality of connections, the level of trust, and the individual's ability to leverage their network position effectively, highlighting the role of quality (McGonigal, 2007). This highlights the need for a more nuanced understanding of how social capital operates in digital environments, moving beyond simple metrics (Park, 2017).

#### 3.2. Symbolic capital and online reputation

The results support Bourdieu's theory of symbolic capital, demonstrating that recognition and prestige are important drivers of social behavior in online environments, validating core concepts (Demidova, NaN). The study confirms that individuals seek to accumulate digital symbolic capital through content creation, social networking, and the cultivation of a positive online reputation (Bourdieu, 2008). This aligns with previous research on the role of symbolic capital in shaping social hierarchies and access to resources, confirming existing theories (Basaran, 2017).

However, the findings also extend Bourdieu's framework by highlighting the unique characteristics of digital symbolic capital, expanding theoretical scope (Huang, 2019). Unlike traditional forms of symbolic capital, digital symbolic capital is more fluid, dynamic, and measurable, enabling tracking and adjustment (Costa, 2013). Online platforms provide tools for tracking visibility, influence, and engagement, allowing individuals to monitor their progress and adjust their strategies accordingly (Singh, 2021).

Moreover, the study reveals that digital symbolic capital is not always aligned with traditional markers of status and prestige, challenging conventional hierarchies (Desrochers, 2018). Individuals who lack formal education or professional credentials can still accumulate significant digital symbolic capital through their online activities, democratizing access (Saburova, 2020). This suggests that digital platforms may offer new pathways for social mobility and recognition, challenging existing power structures (Pokrovskaia, 2017).

### 3.3. Visibility, influence, and power in the digital age

The findings contribute to ongoing debates about the relationship between visibility, influence, and power in the digital age, informing contemporary discussions (Yefanov, 2023). The study confirms that visibility is a key factor in gaining influence, but it also demonstrates that visibility alone is not sufficient, requiring additional factors (Eker, 2015). Credibility, authority, and trust are also essential for shaping attitudes and behaviors in online environments, highlighting the role of authenticity (Brazevich, 2024).

The results align with Castells' theory of the network society, emphasizing the transformative impact of information technologies on power relations, validating existing frameworks (Desrochers, 2018). The study reveals that digital platforms have enabled the creation of new forms of social organization and interaction, challenging traditional hierarchies and empowering marginalized groups (Heidari, 2020). However, the findings also suggest that these new forms of power are not without their limitations and contradictions, raising ethical considerations (Andrade, 2022).

Furthermore, the study highlights the importance of critical analysis in understanding the dynamics of visibility, influence, and power in the digital age, promoting responsible engagement (Ragnedda, 2022). Online platforms can be used to manipulate public opinion, spread misinformation, and promote hate speech, creating risks (Pramanda, 2021). Understanding these negative effects is essential for developing strategies to promote more responsible and ethical communication practices and to protect vulnerable populations from harm, fostering a safer environment (Ragnedda, 2022).

## 4. Conclusion and implications

This study has examined how visibility and influence operate in digital environments to produce a novel form of symbolic capital—digital symbolic capital—that emerges from the interplay of algorithmic structures, user agency, and relational dynamics. Drawing upon the theoretical insights of Bourdieu, Castells, and Cardon, and employing a mixed-methods approach, the findings demonstrate that online recognition is not solely a function of popularity but results from the convergence of content strategy, network positioning, and perceived trust.

- **Theoretical contributions**

This research contributes to the renewal of sociological theory by extending Bourdieu's framework into algorithmically mediated spaces. It refines the notion of symbolic capital by incorporating the digital affordances of visibility and influence. Moreover, it provides empirical grounding to the emerging concept of capital de visibilité (Cardon, 2019), positioning it as a key form of symbolic power in the platform society.

By integrating network theory and digital culture, the study highlights that social stratification online is increasingly shaped by algorithmic logics rather than traditional institutional markers. It thereby challenges the assumptions of meritocracy in digital visibility, showing how symbolic legitimacy is co-constructed by both platform design and user engagement.

- **Practical and political implications**

The findings raise important implications for platform governance, digital inclusion, and influencer practices. First, they suggest the necessity of enhancing algorithmic transparency, as current visibility mechanisms often perpetuate unequal access to attention. Policymakers should consider the introduction of regulatory frameworks to combat algorithmic bias and visibility monopolies, especially when they affect public discourse and democratic participation.

Second, the results underline the importance of developing users' digital literacy, particularly their understanding of engagement algorithms and self-branding strategies. Educational institutions, NGOs, and digital platforms could co-develop training programs that empower marginalized users to participate equitably in the symbolic economy of visibility.

Third, for digital strategists and content creators, the study provides actionable insights on the value of authenticity, emotional resonance, and relational engagement—beyond surface metrics of virality.

- **Limitations and future research**

Like any empirical inquiry, this study is not without limitations. The sample, though analytically rich, remains limited in scope and diversity. Moreover, platform dynamics evolve rapidly, making it difficult to generalize results over time. Future research should adopt longitudinal and cross-platform designs, and explore how digital symbolic capital varies across cultural, gendered, and geopolitical contexts.

In particular, comparative studies between Global North and Global South digital practices could illuminate the asymmetric geographies of digital prestige, offering a more global understanding of visibility dynamics.

In sum, this research advances a critical framework to interrogate how recognition, influence, and power circulate in digital social relations. As platforms continue to structure public visibility, understanding the social architecture behind digital prestige becomes essential-not only for theory-building but also for fostering a more equitable digital society.